\documentclass[prl,aps,floatfix,showpacs,twocolumn]{revtex4}
\usepackage{dcolumn}
\usepackage{bm}
\usepackage{graphicx}
\begin{document}

\title{Tensor Forces and the Ground-State Structure of Nuclei}
\author{R.\ Schiavilla$^{1,2}$, R.B.\ Wiringa$^3$, Steven C.\ Pieper$^3$, and J.\ Carlson$^4$}
\affiliation{
$^1$Jefferson Lab, Newport News, VA 23606 \\
$^2$\mbox{Department of Physics, Old Dominion University, Norfolk, VA 23529}\\
$^3$Physics Division, Argonne National Laboratory, Argonne, IL 61801\\
$^4$\mbox{Theoretical Division, Los Alamos National Laboratory, Los Alamos, NM 87545}
}

\date{\today}

\begin{abstract}
Two-nucleon momentum distributions are calculated for the ground states of
nuclei with mass number $A\leq 8$, using variational Monte Carlo
wave functions derived from a realistic Hamiltonian with two- and three-nucleon
potentials.  The momentum distribution of $np$ pairs is found to be much
larger than that of $pp$ pairs for values of the relative momentum in the
range (300--600) MeV/c and vanishing total momentum.  This
order of magnitude difference is seen in all nuclei considered and has a
universal character originating from the tensor components present in any
realistic nucleon-nucleon potential.  The correlations induced by the tensor
force strongly influence the structure of $np$ pairs, which are
predominantly in deuteron-like states, while they are ineffective for $pp$
pairs, which are mostly in $^1$S$_0$ states.  These features should be easily
observable in two-nucleon knock-out processes, such as $A(e,e^\prime np)$
and $A(e,e^\prime pp)$.
\end{abstract}


\maketitle
The two preeminent features of the nucleon-nucleon ($N\!N$) interaction
are its short-range repulsion and intermediate- to long-range tensor
character.  These induce strong
spatial-spin-isospin $N\!N$ correlations, which leave their imprint on the
structure of ground- and excited-state wave functions.  Several nuclear
properties reflect the presence of these features.
For example, the two-nucleon density distributions
$\rho^{M_S}_{TS}({\bf r})$ in states with pair spin $S$=1 and isospin $T$=0
are very small at small inter-nucleon separation ${\bf r}$ and exhibit strong
anisotropies depending on the spin projection $M_S$~\cite{Forest96}.  Nucleon
momentum distributions $N(k)$~\cite{Schiavilla86,Pieper92} and spectral
functions $S(k,E)$~\cite{Benhar89} have large high-momentum and, in the case of
$S(k,E)$, high-energy components,
which are produced by short-range and tensor correlations.  The
latter also influence the distribution of strength in response functions
$R(k,\omega)$, which characterize the response of the nucleus
to a spin-isospin disturbance injecting momentum ${\bf k}$ and energy $\omega$
into the system~\cite{Fabrocini89,Pandharipande94}.  Lastly, calculations of
low-energy spectra in light nuclei (up to mass number $A$=10) have demonstrated
that tensor forces play a crucial role in reproducing the observed ordering of
the levels and, in particular, the observed absence of stable $A$~=~8
nuclei~\cite{Wiringa02}.

In the present study we show that tensor correlations also impact strongly
the momentum distributions of $N\!N$ pairs in the ground state of a nucleus
and, in particular, that they lead to large differences in the $np$ versus
$pp$ distributions at moderate values of the relative momentum in the pair.
These differences should be observable in two-nucleon knock-out processes,
such as $A(e,e^\prime np)$ and $A(e,e^\prime pp)$ reactions.

The probability of finding two nucleons with relative momentum ${\bf q}$
and total momentum ${\bf Q}$ in isospin state $TM_T$ in the
ground state of a nucleus is proportional to the density
\vspace*{-.2in}
\begin{widetext}
\vspace*{-.2in}
\begin{eqnarray}
\rho_{TM_T}({\bf q},{\bf Q})\!\!&=&\!\!\frac{A(A-1)}{2\, (2J+1)}\! \sum_{M_J}
\int d{\bf r}_1\, d{\bf r}_2\, d{\bf r}_3 \cdots d{\bf r}_A\, 
d{\bf r}^\prime_1\,d{\bf r}^\prime_2 \,
\psi^\dagger_{JM_J}({\bf r}_1^\prime,{\bf r}_2^\prime,{\bf r}_3, \dots,{\bf r}_A)\,  \nonumber \\
&& \times \, e^{-i{\bf q}\cdot ({\bf r}_{12}-{\bf r}_{12}^\prime)} \,
   e^{-i{\bf Q}\cdot ({\bf R}_{12}-{\bf R}_{12}^\prime)}
\, P_{TM_T} (12) \, 
\psi_{JM_J} ({\bf r}_1,{\bf r}_2,{\bf r}_3, \dots,{\bf r}_A) \ ,
\label{eq:rhoqQ}
\end{eqnarray}
\end{widetext}
\vspace*{-.2in}
where 
${\bf r}_{12}\equiv {\bf r}_1-{\bf r}_2$,
${\bf R}_{12}\equiv ({\bf r}_1+{\bf r}_2)/2$, and similarly for
${\bf r}^\prime_{12}$ and ${\bf R}^\prime_{12}$.
$P_{TM_T}(12)$
is the isospin projection operator, and $\psi_{JM_J}$ denotes
the nuclear wave function in spin and spin-projection state
$JM_J$.  The normalization is
\begin{equation}
\int \frac{d{\bf q}}{(2\pi)^3}
 \frac{d{\bf Q}}{(2\pi)^3}\, \rho_{TM_T}({\bf q},{\bf Q})=
N_{TM_T} \ ,
\end{equation}
where $N_{TM_T}$ is the number of $N\!N$ pairs in state
$TM_T$.  
Obviously, integrating $\rho_{TM_T}({\bf q},{\bf Q})$
over only ${\bf Q}$ gives the probability of finding two nucleons with
relative momentum ${\bf q}$, regardless of their pair momentum ${\bf Q}$
(and vice-versa).

The present study of two-nucleon momentum distributions in light nuclei
(up to $A$=8) is based on variational Monte Carlo (VMC) wave functions,
derived from a realistic Hamiltonian consisting of the Argonne $v_{18}$
two-nucleon~\cite{Wiringa95} and Urbana-IX three-nucleon~\cite{Pudliner95}
interactions (AV18/UIX).  The high accuracy of the VMC wave functions is well
documented (see Refs.~\cite{Carlson98,Pieper01} and references therein), as
is the quality of the AV18/UIX Hamiltonian in quantitatively
accounting for a wide variety of light nuclei properties, such as
elastic and inelastic electromagnetic form factors~\cite{Wiringa98},
and low-energy capture reactions~\cite{Marcucci06}.  However, it is important
to stress that the large
effect of tensor correlations on two-nucleon momentum distributions and
the resulting isospin dependence of the latter remain
valid, even if one uses a semi-realistic Hamiltonian model.
This will be shown explicitly below.

The double Fourier transform in Eq.~(\ref{eq:rhoqQ}) is computed by Monte
Carlo (MC) integration.  A standard Metropolis walk, guided by 
$|\psi_{JM_J} ({\bf r}_1,{\bf r}_2,{\bf r}_3, \dots,{\bf r}_A)|^2$, is used
to sample configurations~\cite{Pieper01}.  For each configuration 
a two-dimensional grid of
Gauss-Legendre points, $x_i$ and $X_j$, is used to compute the Fourier transform.
Instead of just moving the $\psi^\prime$ position (${\bf r}_{12}^\prime$ and  
${\bf R}_{12}^\prime$) away from a fixed $\psi$ position (${\bf r}_{12}$ and  
${\bf R}_{12}$), both positions are moved symmetrically away from
${\bf r}_{12}$ and  ${\bf R}_{12}$, so Eq.~(\ref{eq:rhoqQ}) becomes
\vspace*{-.2in}
\begin{widetext}
\vspace*{-.2in}
\begin{eqnarray}
\rho_{TM_T}({\bf q},{\bf Q}) = \frac{A(A-1)}{2\, (2J+1)}\! \sum_{M_J}
\int d{\bf r}_1\, d{\bf r}_2\, d{\bf r}_3 \cdots d{\bf r}_A \, d{\bf x}\, d{\bf X}\, 
\psi^\dagger_{JM_J} 
({\bf r}_{12}\!+\!{\bf x}/2,{\bf R}_{12}\!+\!{\bf X}/2,{\bf r}_3, \dots,{\bf r}_A)&&  \nonumber \\
\times  \,  e^{-i{\bf q}\cdot {\bf x} } \, e^{-i{\bf Q}\cdot {\bf X} } \, P_{TM_T} (12) \,
\psi_{JM_J} 
({\bf r}_{12}\!-\!{\bf x}/2,{\bf R}_{12}\!-\!{\bf X}/2,{\bf r}_3, \dots,{\bf r}_A)&& .
\end{eqnarray}
\end{widetext}
\vspace*{-.2in}
Here the polar angles of the $x$ and $X$ grids are also sampled by MC 
integration, with one sample per pair.
This procedure is similar to that adopted most recently in studies of the
$^3$He$(e,e^\prime p)d$ and $^4$He$(\vec{e},e^\prime \vec{p}\,)^3$H
reactions~\cite{Schiavilla05}, and has the advantage of very substantially
reducing the statistical errors originating from the rapidly oscillating
nature of the integrand for large values of 
$q$ and $Q$.  Indeed, earlier calculations of nucleon 
and cluster momentum 
distributions 
in few-nucleon systems, which were carried out by direct MC
integration over all coordinates, were very noisy 
for momenta beyond 2~fm$^{-1}$, 
even when the random walk consisted of a very large number of
configurations~\cite{Schiavilla86}.

The present method is, however, computationally intensive, 
because complete Gaussian integrations have to be performed for each of
the configurations sampled in the random walk.
The large range of values of $x$ and $X$ required to obtain converged 
results,  especially for $^3$He, require fairly large numbers of points; 
we used grids of up to 96 and 80 points for $x$ and $X$, respectively.
We also sum over all pairs instead of just pair 12.

\begin{figure}[b!]
\includegraphics[angle=-90,width=3.5in]{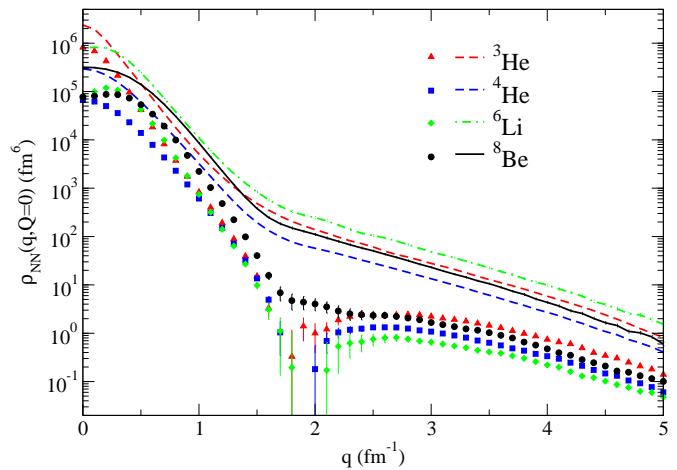}
\caption{(Color online) The $np$ (lines) and $pp$ (symbols) momentum 
distributions in various nuclei as functions of the relative momentum 
$q$ at vanishing total pair momentum $Q$.}
\label{fig:moms}
\end{figure}
The $np$ and $pp$ momentum distributions in $^3$He, $^4$He, $^6$Li, and $^8$Be 
nuclei are shown in Fig.~\ref{fig:moms} as functions of the relative momentum 
$q$ at fixed total pair momentum $Q$=0, corresponding to nucleons moving back
to back.  The statistical errors due to the Monte Carlo integration are 
displayed only for the $pp$ pairs; they are negligibly small for the $np$
pairs.  The striking features seen in all cases are: i) the momentum 
distribution of $np$ pairs is much larger than that of $pp$ pairs for
relative momenta in the range 1.5--3.0 fm$^{-1}$, and ii) for the helium and
lithium isotopes the node in the $pp$ momentum distribution is absent in the 
$np$ one, which instead exhibits a change of slope at a characteristic value of 
$p \simeq 1.5$ fm$^{-1}$.  The nodal structure is much less prominent in 
$^8$Be.  At small values of $q$ the ratios of $np$ to $pp$ momentum 
distributions are closer to those of $np$ to $pp$ pair numbers, which in 
$^3$He, $^4$He, $^6$Li, and $^8$Be are respectively 2, 4, 3, and 8/3.  
Note that the $np$ momentum distribution is given by the linear combination
$\rho_{TM_T=10}+\rho_{TM_T=00}$, while the $pp$ momentum distribution 
corresponds to $\rho_{TM_T=11}$.  The wave functions utilized in the present 
study are eigenstates of total isospin ($1/2$ for $^3$He, and 0 for 
$^4$He, $^6$Li, and $^8$Be), so the small effects of 
isospin-symmetry-breaking interactions are ignored.  
As a result, in $^4$He, $^6$Li, and $^8$Be the $\rho_{TM_T}$ is
independent of the isospin projection and, in particular, the $pp$ and $T=1$
$np$ momentum distributions are the same.

The excess strength in the $np$ momentum distribution is due to the
strong correlations induced by tensor components in the underlying
$N\!N$ potential.  For $Q$=0, the pair and residual $(A$--$2)$
system are in a relative S-wave.  In $^3$He and $^4$He with uncorrelated
wave functions, $3/4$ of the $n$$p$ pairs are in deuteron-like
$T,S$=0,1 states, while the $p$$p$, $nn$ and remaining $1/4$
of $n$$p$ pairs are in $T,S$=1,0 (quasi-bound) states.  When multi-body 
tensor correlations are taken into account, 10--15\% of the $T,S$=1,0 pairs 
are spin-flipped to $T,S$=1,1 pairs, but the number of $T,S$=0,1 pairs hardly 
changes~\cite{Forest96}.
In $A>4$ nuclei, some $n$$p$ and $p$$p$ pairs will be in relative P-waves
($T,S$=1,1 and 0,0) when one particle is in the s-shell and one in the
p-shell.  Nevertheless, 5.5 out of 9 $n$$p$ pairs in $^6$Li (9 out of 16 in
$^8$Be) are expected to be deuteron-like, while half the $p$$p$ pairs will be
in $T,S$=1,0 states and half in $T,S$=1,1 states~\cite{Wiringa06}.  
The tensor force vanishes in the former and is weak in the latter.

\begin{figure}[b!]
\includegraphics[angle=-90,width=3.5in]{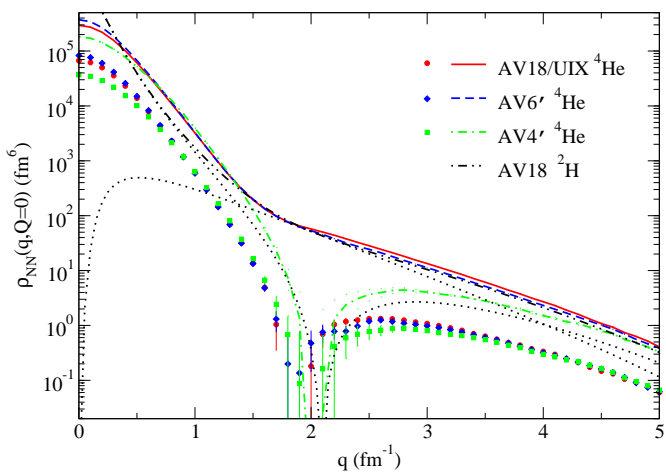}
\caption{(Color online) The $np$ (lines) and $pp$ (symbols) momentum 
distributions in $^4$He obtained with different Hamiltonians.  Also shown 
is the scaled momentum distribution for the AV18 deuteron; its separate 
S- and D-wave components are shown by dotted lines.}
\label{fig:cmp}
\end{figure}
These expectations are born out by our calculations,
as Fig.~\ref{fig:moms} clearly demonstrates.  The $np$ momentum distributions
for $q$ values larger than 1.5 fm$^{-1}$ only differ by a scaling factor, 
and indeed all scale relative to the deuteron momentum distribution as 
shown in Fig.~\ref{fig:cmp}.  The deuteron $\rho_{np}(q,Q=0)$
has been scaled to match that of $^4$He at $q=1.5$~fm$^{-1}$.
(The scaling property does not extend to the low $q$ region, because
there binding effects take over.)  The S- and D-wave components of the
deuteron density are also shown in the figure as dotted lines; the D-state
component is the dominant part over the range 1.4--4.0 fm$^{-1}$, while
the S-state component has a node at 2.1 fm$^{-1}$.

Similar considerations in relation to scaling and binding effects also remain
valid when considering the $pp$ momentum distributions.  In particular, the
node seen in helium and lithium is reminiscent of the node in the S-wave
momentum distribution (shown in Fig.~\ref{fig:cmp}).  In larger nuclei
the node is filled in, due to the fact that $p$$p$ pairs are not exclusively in
quasi-bound states, but can also be in P-wave or higher partial-wave states.

Figure~\ref{fig:cmp} also shows the $n$$p$ and $p$$p$ momentum distributions
in $^4$He obtained with Hamiltonians of decreasing sophistication,
ranging from the fully realistic AV18/UIX model to the semi-realistic 
Argonne $v_6^\prime$ (AV6$^\prime$) to the relatively simple Argonne 
$v_4^\prime$ (AV4$^\prime$).  The AV4$^\prime$ and AV6$^\prime$
potentials are constructed to preserve as many features of $N\!N$
scattering and deuteron properties as feasible~\cite{Wiringa02}.  The
AV4$^\prime$ has only central, spin, and isospin operators with no tensor 
component: it reproduces the $^1$S$_0$ phase shift, and the deuteron
binding energy but with only an S-state component; the D-state, 
induced by the tensor term in the potential and associated with one-pion 
exchange at long range, is absent.  On the other hand, the AV6$^\prime$ 
includes tensor terms and produces a bound deuteron quite close to that of AV18.
These features of the underlying $N\!N$ potential, or the lack of
them, are reflected in the calculated momentum distributions.  In 
particular, note the node which develops in the AV4$^\prime$ $np$ momentum
distribution, due to the purely S-wave nature of the deuteron-like state.
On the other hand, the AV6$^\prime$ and AV18/UIX results are very
close to each other, demonstrating the essential role played by the 
tensor potential in substantially increasing the intermediate-momentum components of
$np$ pairs.  

\begin{figure}[bth]
\includegraphics[angle=-90,width=3.5in]{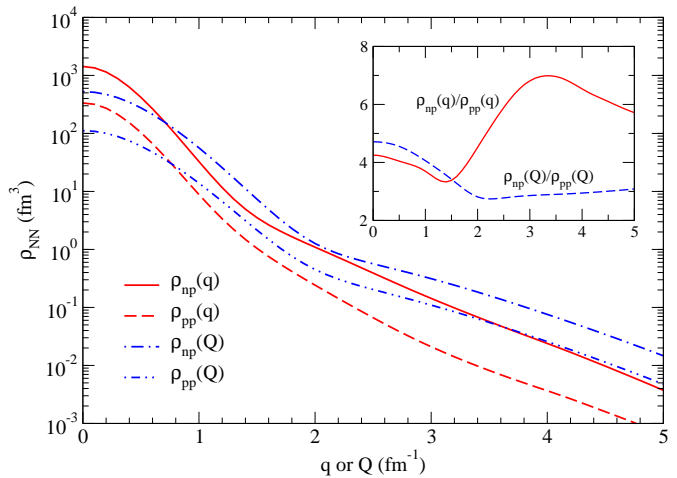}
\caption{(Color online) The momentum distributions $\rho_{NN}(q)$ and
$\rho_{NN}(Q)$ in $^4$He for $np$ and $pp$.  The inset shows the
ratios $\rho_{np}(q)/\rho_{pp}(q)$ and $\rho_{np}(Q)/\rho_{pp}(Q)$.}
\label{fig:pmoms}
\end{figure}
The momentum distributions $\rho_{NN}(q)$ and $\rho_{NN}(Q)$ obtained by
integrating over $Q$ or $q$ respectively
are plotted in Fig.~\ref{fig:pmoms} for $^4$He with 
the AV18/UIX Hamiltonian.  There is no node in
$\rho_{pp}(q)$: it is filled in by the contributions of $p$$p$ channels
other than $^1$S$_0$.  These channels are now allowed, because the orbital
angular momentum between the $p$$p$ pair and the residual $n$$n$ cluster
is not constrained to vanish for $Q>0$.  A remnant of the strong tensor
correlations affecting $\rho_{np}(q)$ still persists.  This is illustrated in
the inset of Fig.~\ref{fig:pmoms}, which shows the ratio $\rho_{np}(q)/\rho_{pp}(q)$.
However, the effect is far less dramatic
than in the back-to-back ($Q$=0) kinematics.  Note that $\rho_{NN}(q)$ and
$\rho_{NN}(Q)$ have the same normalization, {\it i.e.} $(A-Z)Z$ for $n$$p$
and $Z(Z-1)/2$ for $p$$p$.  The $\rho_{np}(Q)/\rho_{pp}(Q)$ ratio is close
to the $np$ to $pp$ pair number ratio---4 in $^4$He---over the whole
range of $Q$ values.  This holds true for all the nuclei studied.

The most direct evidence for tensor
correlations in nuclei comes from measurements of the
deuteron structure functions and tensor polarization by elastic electron
scattering~\cite{Alexa99}.  In essence, these measurements have mapped out the
Fourier transforms of the charge densities of the deuteron in states with spin
projections $\pm1$ and $0$, showing that they are very different.  In other
processes, such as $^2$H$(d,\gamma)^4$He~\cite{Arriaga91} 
at very low energy, or proton knock-out from a
polarized deuteron~\cite{Zhou99} (as well as in the nuclear properties
mentioned at the beginning of this letter), the effects of tensor
correlations are more subtle and their presence is not easily isolated in
the experimental data.  This is because of ``contaminations'' from initial or
final state interactions and many-body terms in the transition operators.

Some of these corrections will also pollute the cross sections for 
$(e,e^\prime np)$ and $(e,e^\prime pp)$, or $(p,pp)$ and $(p,ppn)$,
knock-out processes in back-to-back kinematics.  However, one would expect
the contributions due to final state interactions in the $np$ and $pp$
reactions, both between the nucleons in the pair and between these and
the nucleons in the residual $(A-2)$ system, to be of similar magnitude
for relative momenta in the range (300--600) MeV/c.  In the electrodisintegration 
processes, the leading electromagnetic two-body currents associated with
pion and $\rho$-meson exchange, denoted respectively as $PS$ and $V$ in
Ref.~\cite{Marcucci05}, vanish in $pp$ because of their isospin structure.
Of course, they will contribute in $np$, but are not expected to produce
large effects.  Thus the ratio of $np$ to $pp$ cross sections should be much
larger than unity for relative momenta within (300--600) MeV/c, reflecting the
large difference between the corresponding momentum distributions in this range.
There are strong indications from a recent analysis of a BNL experiment,
which measured cross sections for $(p,pp)$ and $(p,ppn)$ processes
on $^{12}$C in kinematics close to two nucleons being ejected back to back,
that this is indeed the case~\cite{Piasetzky06a}.  The ratio of $(p,ppn)$ to
$(p,pp)$ events over the range of relative momenta (275--550) MeV/c is found
to be roughly 20, albeit with a rather large error.  Hopefully, a more
precise value for this ratio will become available in the near future,
when the analysis of $^{12}$C$(e,e^\prime np)$ and
$^{12}$C$(e,e^\prime pp)$ data, taken at Jefferson Lab,
is completed~\cite{Piasetzky06b}.

It would be interesting to extend these measurements to other nuclei.
In $^3$He and $^4$He, one would expect the node in the $pp$ momentum
distribution to be filled in by interaction effects in the final
state~\cite{Schiavilla05}.
However, the ratio of $np$ to $pp$ cross sections in the range (300--600)
MeV/c should still reflect the large value of the $np$ momentum
distribution at these values of relative momenta.  This would provide a
further, direct verification of the crucial role that the tensor force
plays in shaping the short-range structure of nuclei.

We dedicate this paper to the memory of Vijay R. Pandharipande,
a mentor and friend, who was deeply interested in evidence for
correlations in nuclei.  This is also in remembrance of our
colleague and friend Adelchi Fabrocini, who contributed significantly
to the theoretical study of correlations in strongly interacting
systems.
An interesting and stimulating conversation with D.\ Higinbotham and
E.\ Piasetzky is gratefully acknowledged by one of the authors (R.S.).
This work is supported by the U.S.\ Department of Energy, Office of
Nuclear Physics, under contracts
DE-AC05-06OR23177 (R.S.), DE-AC-02-06CH11357 (S.C.P.\ and R.B.W.), and 
W-7405-ENG-36 (J.C.).  The calculations were made at Argonne's Laboratory 
Computing Resource Center.

\begin{thebibliography}{100}
%
%
\bibitem{Forest96} J.\ L.\ Forest {\it et al.},
Phys.\ Rev.\ C {\bf 54}, 646 (1996).
%
\bibitem{Schiavilla86} R.\ Schiavilla, V.\ R.\ Pandharipande, and R.\ B.\ Wiringa,
Nucl.\ Phys.\ A {\bf 449}, 219 (1986).
%
\bibitem{Pieper92} S.\ C.\ Pieper, R.\ B.\ Wiringa, and V.\ R.\ Pandharipande,
Phys.\ Rev.\ C {\bf 46}, 1741 (1992).
%
\bibitem{Benhar89} O.\ Benhar, A.\ Fabrocini, and S.\ Fantoni,
Nucl.\ Phys.\ A {\bf 505}, 267 (1989).
%
\bibitem{Fabrocini89} A.\ Fabrocini and S.\ Fantoni,
Nucl.\ Phys.\ A {\bf 503}, 375 (1989).
%
\bibitem{Pandharipande94} V.\ R.\ Pandharipande {\it et al.},
Phys.\ Rev.\ C {\bf 49}, 789 (1994).
%
\bibitem{Wiringa02} R.\ B.\ Wiringa and S.\ C.\ Pieper,
Phys.\ Rev.\ Lett.\ {\bf 89}, 182501 (2002).
%
\bibitem{Wiringa95}
R.\ B.\ Wiringa, V.\ G.\ J.\ Stoks, and R.\ Schiavilla,
Phys.\ Rev.\ C {\bf 51}, 38 (1995).
%
\bibitem{Pudliner95} B.\ S.\ Pudliner {\it et al.},
Phys.\ Rev.\ Lett.\ {\bf 74}, 4396 (1995).
%
\bibitem{Carlson98} J.\ Carlson and R.\ Schiavilla,
Rev.\ Mod.\ Phys.\ {\bf 70}, 743 (1998).
%
\bibitem{Pieper01} S.\ C.\ Pieper and R.\ B.\ Wiringa,
Annu.\ Rev.\ Nucl.\ Part.\ Sci.\ {\bf 51}, 53 (2001).
%
\bibitem{Wiringa98} R.\ B.\ Wiringa and R.\ Schiavilla,
Phys.\ Rev.\ Lett.\ {\bf 81}, 4317 (1998);
L.\ E.\ Marcucci, D.\ O.\ Riska, and R.\ Schiavilla,
Phys.\ Rev.\ C {\bf 58}, 3069 (1998).
%
\bibitem{Marcucci06} L.\ E.\ Marcucci {\it et al.},
Nucl.\ Phys.\ A {\bf 777}, 111 (2006).
%
\bibitem{Schiavilla05} R.\ Schiavilla {\it et al.},
Phys.\ Rev.\ Lett.\ {\bf 94}, 072303 (2005); Phys.\ Rev.\ C {\bf 72}, 064003 (2005).
%
\bibitem{Wiringa06} R.\ B.\ Wiringa,
Phys.\ Rev.\ C {\bf 73}, 034317 (2006).
%
\bibitem{Alexa99} A complete list of references is in R.\ Gilman and F.\ Gross,
J.\ Phys.\ {\bf G28},  R37 (2002).
%
\bibitem{Arriaga91} A.\ Arriaga, V.\ R.\ Pandharipande, and R.\ Schiavilla,
Phys.\ Rev.\ C {\bf 43}, 983 (1991).
%
\bibitem{Zhou99} Z.-L.\ Zhou {\it et al.},
Phys.\ Rev.\ Lett.\ {\bf 82}, 687 (1999);
I.\ Passchier {\it et al.},
Phys.\ Rev.\ Lett.\ {\bf 88}, 102302 (2002).
%
\bibitem{Marcucci05} L.\ E.\ Marcucci {\it et al.},
Phys.\ Rev.\ C {\bf 72}, 014001 (2005).
%
\bibitem{Piasetzky06a} E.\ Piasetzky {\it et al.},
Phys.\ Rev.\ Lett.\ {\bf 97}, 162504 (2006).
%
\bibitem{Piasetzky06b} E.\ Piasetzky and D.\ Higinbotham,
private communication.
%
%
\end{thebibliography}
\end{document}